\documentclass[aps,twocolumn,prl,floats,amsmath,amssymb,showpacs]{revtex4-1}
\RequirePackage{xspace} 
\usepackage[bookmarks,colorlinks]{hyperref}
\usepackage{epsfig}
\usepackage{graphicx}
\usepackage{color}
\usepackage{mathrsfs}
\usepackage{amssymb}
\usepackage{amsmath}
\usepackage{url}
\usepackage{nicefrac}

\def\dsp{\displaystyle}
\def\be {\begin{equation}}
\def\ee {\end{equation}}
\def\bea {\begin{eqnarray}}
\def\eea {\end{eqnarray}}
\def\bc {\begin{center}}
\def\ec {\end{center}}
\def\nn {\nonumber}

\def\AFB{A_{\text{FB}}}
\def\Gf{\Gamma_{\!\! f}}
\def \thl {{\theta_\ell}}
\def \thK {{\theta_{K}}}
\def \azeL{{{\cal A}_0^L}}

\def \apaL{{{\cal A}_\parallel^L}}

\def \apeL{{{\cal A}_\perp^L}}

\def \Re{\text{Re}}

\def \kstar{{K^{\!*}}}
\def \eff{{\text{eff }}}

\newcommand{\tev}{\ensuremath{\mathrm{Te\kern -0.1em V}}\xspace}
\newcommand{\gev}{\ensuremath{\mathrm{Ge\kern -0.1em V}}\xspace}
\newcommand{\mev}{\ensuremath{\mathrm{Me\kern -0.1em V}}\xspace}
\newcommand{\kev}{\ensuremath{\mathrm{ke\kern -0.1em V}}\xspace}
\newcommand{\ev}{\ensuremath{\mathrm{e\kern -0.1em V}}\xspace}
\newcommand{\gevc}{\ensuremath{{\mathrm{Ge\kern -0.1em V\!/}c}}\xspace}
\newcommand{\mevc}{\ensuremath{{\mathrm{Me\kern -0.1em V\!/}c}}\xspace}
\newcommand{\gevcc}{\ensuremath{{\mathrm{Ge\kern -0.1em V\!/}c^2}}\xspace}
\newcommand{\mevcc}{\ensuremath{{\mathrm{Me\kern -0.1em V\!/}c^2}}\xspace}

\def\ket|#1>{\left|#1 \right>}

\def\bra<#1|{\left< #1 \right|}

\def\bracket<#1|#2>{\setbox0=\vbox{\hbox{$#1$$#2$}}\left<#1\kern1pt
    \vrule  height\ht0\kern2pt #2\right>} 

\def\dirmat<#1|#2|#3>{\setbox0=\vbox{\hbox{$#1$$#2$$#3$}}\left<#1\kern1pt
    \vrule height\ht0\kern1pt#2\kern1pt \vrule height\ht0\kern1pt
    #3\right>}

\allowdisplaybreaks
\begin{document}

\title{Isolating New Physics Effects from Hadronic Form Factor
  Uncertainties in $\mathbf{B\to K^* \ell^+\ell^-}$}

\author{Diganta Das}
\affiliation{The Institute of Mathematical Sciences, Taramani, Chennai
  600113, India}
\author{Rahul Sinha}
\affiliation{The Institute of Mathematical Sciences, Taramani, Chennai
  600113, India}

\date{ \today}

%
%
\begin{abstract}
  The discovery of New Physics, using weak decays of mesons is
  difficult due to intractable strong interaction effects needed to
  describe it. We show how the multitude of ``related observables''
  obtained from $B\to \kstar \ell^+\ell^-$, can provide many new
  ``clean tests'' of the Standard Model. The hallmark of these tests
  is that several of them are independent of the unknown form factors
  required to describe the decay using heavy quark effective theory.
  We derive a relation between observables that is free of form
  factors and Wilson coefficients, the violation of which will be an
  unambiguous signal of New Physics. We also derive other relations
  between observables and form factors that are independent of Wilson
  coefficients and enable verification of hadronic estimates.  We find
  that the allowed parameter space for observables is very tightly
  constrained in Standard Model, thereby providing clean signals of
  New Physics. The relations derived will provide unambiguous signals
  of New Physics if it contributes to these
  decays.
\end{abstract}


\maketitle

Indirect searches for New Physics often involve precision measurement
of a single quantity.  The most well known example is the muon
magnetic moment. Unfortunately, even though muon is a lepton,
estimating hadronic contributions turn out to be the limiting factor
in the search for New Physics.  We show how certain B meson decays
offer instead a multitude of related observables that enable handling
intractable strong interaction effects that hamper the discovery of
New Physics. It is hoped that flavor changing neutral current
transitions of the $b$ quark will be altered by physics beyond the
Standard Model (SM) and their study would reveal possible signal of
New Physics (NP) if it exists. However, understanding the hadronic
flavor changing neutral current decays requires estimating hadronic
effects that are difficult to compute accurately. Experiments seem to
indicate that new physics does not show up as a large and unambiguous
effect in flavor physics. This has bought into focus the need for
theoretically cleaner observables, i.e. observables that are
relatively free from hadronic uncertainties. In the search for new
physics, it is therefore crucial to effectively isolate the effect of
new physics from hadronic uncertainties that contribute to the decay.

One of the mode that is regarded as significant in this attempt is
$B\to \kstar \ell^+\ell^-$, an angular analysis of which is known to
result in a multitude of observables \cite{Sinha:1996sv,
  Kruger:1999xa}, that enable testing the SM and probing possible NP
contributions~\cite{Kruger:1999xa, Altmannshofer:2008dz,
  Bobeth:2008ij, Egede:2008uy, Bobeth:2010wg}. Experimentally, angular
analysis of $B\to \kstar \ell^+\ell^-$ decays has already been studied
and a few of the observables already
measured~\cite{:2008ju,:2009zv,Aaltonen:2011cn,Aaij:2011aa}.
The decay $\bar{B}(p)\to K^{*}(k)\ell^+(q_1)\ell^-(q_2)$ with
$K^*(k)\to K(k_1)\pi(k_2)$ on the mass shell, is completely described
by four independent kinematic variables. These are the lepton-pair
invariant mass squared $q^2=(q_1+q_2)^2$, the angle $\phi$ between the
decay planes formed by $\ell^+\ell^-$ and $K\pi$ respectively and the
angles $\theta_K$ and $\theta_\ell$ of the $K$ and $\ell^-$
respectively with the $z$-axis defined in the respective rest frames
of $K^*$ and $\ell^+\ell^-$, assuming that $K^*$ has momentum along
the $+\hat{z}$ in $B$ rest frame. The differential decay rate is,
\begin{align}
  &\frac{d^4\Gamma(B\to K^*\ell^+\ell^-)}{dq^2\, d^3\Omega} =
   \frac{9}{32\pi}\Big[ (I_1^s +I_2^s \cos 2\thl)\sin^2\thK \nn\\&~
   + ( I_1^c + I_2^c  \cos 2\thl)\cos^2\thK
 + I_3 \sin^2\thK \sin^2\thl \cos 2\phi  \nn \\ &~ + I_4 \sin
  2\thK \sin 2\thl \cos\phi + I_5 \sin 2\thK \sin\thl \cos\phi \nn\\&~+ I_6^s
  \sin^2\thK \cos\thl+ I_7 \sin 2\thK \sin\thl \sin\phi\nonumber \\ &~ 
  + I_8 \sin 2\thK \sin 2\thl \sin\phi\! +\! I_9 \sin^2\thK \sin^2\thl
  \sin 2\phi\Big].
\end{align}
where, $d^3\Omega=d\cos\thl d\cos\thK d\phi$ and $I^{(c,s)}_i$ are the
coefficients that can be easily measured by studying the angular
distribution. In the absence of CP-violation the conjugate mode
$\bar{B}\to\bar{K}^*\ell^+\ell^-$ has an identical decay distribution
except that $I_{5,6,8,9} \to -I_{5,6,8,9}$ in the differential decay
distribution as a consequence of $CP$
properties~\cite{Kruger:1999xa}. In our discussions we neglect the
lepton and $s$-quark masses. We also ignore the very small CP
violation arising within the SM and exclude studying the resonant
region. Under these approximations all the Wilson coefficients and
form factors contributing to the decay are real. This results in only
six of the $I$'s being non-zero and independent, allowing only six
independent observables to be measured. Any observable that is chosen
may eventually be expressed in terms of six real transversity
amplitudes ${\cal A}_\lambda^{L,R}$, where $L,R$ denote the chirality
of the lepton $\ell^-$ and $\lambda=\{\perp,\parallel,0\}$ are the
helicity that contribute. The six observables we choose are the
helicity fractions and angular asymmetries that can be easily measured
by studying the angular distributions.  We define the observables
$F_\lambda=(|{\cal A}_\lambda^L|^2+|{\cal
  A}_\lambda^R|^2)/\Gamma_{\!\! f}$, $ \Gamma_{\!\!
  f}\equiv\sum_\lambda(|{\cal A}_\lambda^L|^2+|{\cal
  A}_\lambda^R|^2)$. The longitudinal helicity fraction $F_0$ is often
referred to as $F_L$ in literature and we henceforth use $F_L$ to
denote the longitudinal helicity fraction.  The forward--backward
asymmetry is defined as \cite{note1}
\begin{align}
  \label{eq:A_FB_1}
  A_{\text FB}=\frac{\int_D d\cos\thl \frac{d^2
      (\Gamma+\bar{\Gamma})}{d q^2d\cos\thl}}{\int_{-1}^1d\cos\thl
    \frac{d^2 (\Gamma+\bar{\Gamma})}{d q^2d\cos\thl}}~,
\end{align}
where $\int_D\equiv\int_0^1-\int_{-1}^0$.
Two more asymmetries can be defined as follows:
\begin{align}
  \label{eq:A_4}
  A_{4}=\frac{\int_{D_{LR}}d\phi\int_D d\cos\thK\int_D d\cos\thl
    \frac{d^4(\Gamma-\bar{\Gamma})}{dq^2d^3\Omega}} 
  {\int_0^{2\pi}d\phi\int_{-1}^1d\cos\thK \int_{-1}^1d\cos\thl\,
    \frac{d^4(\Gamma+\bar{\Gamma})}{dq^2d^3\Omega}}\\
  \label{eq:A_5}
  A_{5}=\frac{\int_{-1}^1 d\cos\thl\int_{D_{LR}}d\phi \int_D d\cos\thK
    \frac{d^4(\Gamma+\bar{\Gamma})}{dq^2d^3\Omega} }
  {\int_{-1}^1d\cos\thl\int_0^{2\pi}d\phi\int_{-1}^1d\cos\thK \, 
\frac{d^4(\Gamma+\bar{\Gamma})}{dq^2d^3\Omega}}
\end{align}
where
$\int_{D_{LR}}\equiv\int_{\pi/2}^{3\pi/2}-\int_{-\pi/2}^{\pi/2}$. $A_4$,
$A_5$ and $\AFB$ isolate terms proportional to $I_4$, $I_5$ and $I_6$
respectively.  The $I$'s are expressed in terms of amplitudes ${\cal
  A}_{\perp,\parallel,0}^{L,R}$ as:
\begin{align}
  I_1^s &=I_2^s = \frac{3}{4} \big[|\apeL|^2 + |\apaL|^2 +
    (L\to R) \big], \nn\\ 
  I_1^c &=-I_2^c = \big[|\azeL|^2 + (L\to R)\big],\nn\\
  I_3 & = \tfrac{1}{2}\big[ |\apeL|^2 - |\apaL|^2  + (L\to
    R)\big],\nn\\
  I_4 & = \tfrac{1}{\sqrt{2}}\big[\Re (\azeL^{}\apaL^*) +
    (L\to R)\big],\nn\\
  I_5 & = \sqrt{2}\big[\Re(\azeL^{}\apeL^*) - (L\to R)
\big]\nn\\
  I_6^s  & = 2\big[\Re (\apaL^{}\apeL^*) - (L\to R) \big].\nn
\end{align}
We note that even though the helicity amplitudes ${\cal
  A}_\lambda^{L,R}$ are functions of $q^2$, for simplicity we have
suppressed the explicit dependence on $q^2$.

The six transversity amplitudes can be written in the most general
form  as follows:
\begin{equation}
\label{strtg-def1}
  \mathcal{A}_\lambda^{ L,R}=C_{L,R}\mathcal{F}_\lambda-
  \widetilde{\mathcal{G}}_\lambda 
\end{equation}
where to leading order, $C_{L,R}=C_9^\eff\mp C_{10}$ and
$\widetilde{\mathcal{G}}_\lambda= C_7^\eff
\mathcal{G}_\lambda$. $C_7^\eff$, $C_9^\eff$ and $C_{10}$ are the
Wilson coefficients that represent short distance corrections.
$\mathcal{F}_\lambda$ and $\widetilde{\mathcal{G}}_\lambda$ are
defined~\cite{Das-Sinha-II} in terms of $q^2$-dependent QCD form
factors that parameterize the $B\to \kstar$ matrix
element~\cite{Altmannshofer:2008dz} and are suitably defined to
include both factorizable and non-factorizable contributions at any
given order.  The hadronic form factors have been calculated using QCD
sum rules on the light cone and in the heavy quark limit using QCD
factorization~\cite{Beneke:2001at} and soft-collinear
theory~\cite{SCET1} that is valid for small $q^2$ (large recoil of
$\kstar$) and using operator product expansion~\cite{Grinstein:2004vb}
which is valid for large $q^2$ (low recoil). In this letter we limit
our discussion to large recoil region for numerical estimates. This is
sufficient to demonstrate the merits of our approach, even though our
approach is valid for all $q^2$~\cite{Das-Sinha-II}.  In the large
recoil limit the next to leading order effects can be parametrically
included by replacements $C_9^\eff\to C_9$ and defining
$\widetilde{\mathcal{G}}_\lambda= C_7^\eff
\mathcal{G}_\lambda+\cdots$, with the dots representing the next to
leading and higher order terms.  We note that even at leading order it
is impossible to distinguish between $C_7^\eff$ and
$\mathcal{G}_\lambda$. Hence the Wilson coefficient and form factor
can be lumped together into a single factor $\widetilde{\cal
  G}_\lambda$. We also emphasize that our analytic conclusions do not
depend (for most of the part) on the explicit expression of the form
factors.

The six observables or equivalently the amplitudes
$\mathcal{A}_\lambda^{ L,R}$ are cast in Eq.~(\ref{strtg-def1}) in
terms of six form factors $\mathcal{F}_\lambda$,
$\widetilde{\mathcal{G}}_\lambda$ and two Wilson coefficients $C_9$
and $C_{10}$. If two additional inputs can be used all the parameters
can be solved in terms of observables.  Fortunately, advances in our
understanding of these form-factors permit us to make two reliable
inputs in terms of ratios of form-factors
${\mathcal{F}_\perp}/{\mathcal{F}_\|}$ and
${\mathcal{G}_\perp}/{\mathcal{G}_\|}$ which are well predicted at
next to leading order in QCD corrections and free from form factors
$\xi_\|$ and $\xi_\perp$ in heavy quark effective theory.  In this
letter we will make an additional assumption of the ratio $R\equiv
C_9/C_{10}$ which has also been reliably calculated in the standard
model; this allows us to make predictions of yet unmeasured
observables.  Deviations of these observables from our predicted
values would be an unambiguous sign of new physics or a failure of
understanding hadronic effects that are considered most reliably
estimated. We derive several important relations between observables,
Wilson coefficients and form factors.  We find that the six
observables are not independent as there exists one constraint relation
that involves observables alone. 
A relation that is derived based
entirely on the assumption that the amplitudes have the form in
Eq.~\eqref{strtg-def1}, but which is never-the-less independent of
form factors and Wilson coefficients would provide an unambiguous test
of the standard model relying purely on observables. We present such a
relation as a one of the central results in this letter and
Ref.~\cite{Das-Sinha-II}.

We begin by considering only three of the observables mentioned
above which can be expressed in our notation as follows:
\begin{align}
\label{eq:F_parallel}
F_\|\Gamma_{\!\! f}=& 2(C_9^2+C_{10}^2)\mathcal{F}_\|^2+
2\widetilde{\mathcal{G}}_\|^2-4C_9\mathcal{F}_\|\widetilde{\mathcal{G}}_\| \\
\label{eq:F_perp}
F_\perp\Gamma_{\!\! f}=&2(C_9^2+C_{10}^2)\mathcal{F}_\perp^2+ 2
\widetilde{\mathcal{G}}_\perp^2-4C_9\mathcal{F}_\perp\widetilde{\mathcal{G}}_\perp
\\ 
\label{eq:AFB}
A_{\text FB}\Gamma_{\!\! f}=&\,3 C_{10}(\mathcal{F}_\|\widetilde{\mathcal{G}}_\perp
+\mathcal{F}_\perp\widetilde{\mathcal{G}}_\|)-6C_9C_{10} \mathcal{F}_\|\mathcal{F}_\perp~.
\end{align}
The solution to the Wilson coefficients is easily obtained by defining
 intermediate variables
$r_\lambda=\dsp\nicefrac{\dsp\widetilde{\mathcal{G}}_\lambda}{\dsp\mathcal{F}_\lambda}-C_9$. We briefly
sketch the procedure: Eqs.~(\ref{eq:F_parallel}), (\ref{eq:F_perp})
and (\ref{eq:AFB}) get written as 
\begin{align}
\label{eq:F1_parallel}
F_\|\Gamma_{\!\! f}=& 2\mathcal{F}_\|^2(r_\|^2+C_{10}^2)\\
\label{eq:F1_perp}
F_\perp\Gamma_{\!\! f}=& 2\mathcal{F}_\perp^2(r_\perp^2+C_{10}^2)\\
\label{eq:1AFB}
A_{\text FB}\Gamma_{\!\! f}=&3\,C_{10}
\mathcal{F}_\|\mathcal{F}_\perp(r_\|+r_\perp) ~. 
\end{align}
We next define ratios of form factors $\mathsf{P_1}$ and $\mathsf{P'_1}$ as follows:
\begin{align}
  \label{eq:3}
  \mathsf{P_1}=&\frac{\mathcal{F}_\perp}{\mathcal{F}_\|}=
  \frac{-\sqrt{{\lambda}(m_B^2,
      m_\kstar^2,q^2)}}{(m_B+m_\kstar)^2}\frac{V(q^2)}{A_1(q^2)}\\
  \mathsf{P'_1}=&\frac{\widetilde{\mathcal{G}}_\perp}{\widetilde{\mathcal{G}}_\|}=
  \frac{-\sqrt{{\lambda}(m_B^2,
      m_\kstar^2,q^2)}}{m_B^2-m_\kstar^2}\frac{\mathcal{T}_1(q^2)}
  {\mathcal{T}_2(q^2)}   
\end{align}
where $M_B$ and $m_\kstar$ are the masses of the $B$ meson and
$\kstar$ meson respectively; $\lambda(a,b,c)= a^2 + b^2 + c^2 - 2 (a
b+ b c + a c)$. $V(q^2)$, $A_1(q^2)$, $\mathcal{T}_1(q^2)$ and
$\mathcal{T}_2(q^2)$ are the well known $B\to \kstar$ hadronic
form-factors defined in Ref.~\cite{Beneke:2001at}, where the complete
expressions for the ``effective photonic form-factors''
$\mathcal{T}_{1,2}$ up to NLO are also given. In the large recoil limit
the form factors reduce to the simple form
$A_1(q^2)/V(q^2)=2Em_B/(m_B+m_\kstar)^2$ and
$\mathcal{T}_1(q^2)/\mathcal{T}_2(q^2)=m_B/(2E)$. Notice that
$\mathsf{P_1}$ and $\mathsf{P'_1}$ are independent of universal form
factors $\xi_\|(q^2)$ and $\xi_\perp(q^2)$ in the effective theory.

One easily solves for $r_\|+r_\perp$ to be,
\begin{equation*}
 \label{eq:4}
  r_\|+r_\perp=\frac{\pm\sqrt{\Gf}}{\sqrt{2} \mathcal{F}_\|}\!\Bigg[\mathsf{P_1^2}
  F_\|+F_\perp\pm \mathsf{P_1} \sqrt{4 F_\|F_\perp-\tfrac{16}{9} A_{\text
      FB}^2}\Bigg]^{\frac{1}{2}} 
\end{equation*}
Notice Eq.~(\ref{eq:1AFB}) implies that for $A_{\text FB}=0$, one must
have $r_\|+r_\perp=0$, which reduces the sign ambiguity in the
solution of $r_\|+r_\perp$. The choice of the sign inside the square
bracket is determined by the fact that $\mathsf{P_1}$ is negative. In fact, for
$A_{\text{FB}}=0$, the vanishing of the square bracket implies that we
must have the exact equality,
\begin{equation}
  \label{eq:P_1 exp}
\mathsf{P_1}=-\frac{\sqrt{F_\perp}}{\sqrt{F_\|}}\Bigg|_{\AFB=0}
\end{equation}
enabling a measurement of $\mathsf{P_1}$ in terms of the ratio of helicity
fractions. If zero crossing were to occur it would provide an
interesting test of our understanding of form
factors.

 It is now easy to derive the
following relations for $C_9$, $C_{10}$ and $\widetilde{\mathcal{G}}_\|$:
\begin{align}
  \label{eq:C9}
  C_9=& \frac{\sqrt{\Gamma_{\!\! f}}}{\sqrt{2}\mathcal{F}_\|}\frac{(F_\|\mathsf{P_1} \mathsf{P'_1}-F_\perp) -\tfrac{1}{2}(\mathsf{P_1}-\mathsf{P'_1})
    Z_1}{(\mathsf{P_1}-\mathsf{P'_1})\bigg[\pm\sqrt{\mathsf{P_1^2} F_\|+F_\perp+ \mathsf{P_1}
    Z_1}\bigg]},\\
    \label{eq:C10}
  C_{10}=&\frac{\sqrt{\Gamma_{\!\! f}}}{\sqrt{2}\mathcal{F}_\|}\frac{2}{3}\frac{
    A_{\text FB}}{\bigg[\pm\sqrt{\mathsf{P_1^2} F_\|+F_\perp+ \mathsf{P_1}
    Z_1}\bigg]}~,\\
  \label{eq:C7}
  \widetilde{\mathcal{G}}_\|=&\frac{\sqrt{\Gamma_{\!\! f}}}{\sqrt{2}}\frac{(\mathsf{P_1^2}
    F_{\|}-F_\perp)}{(\mathsf{P_1}-\mathsf{P'_1})\bigg[\pm\sqrt{\mathsf{P_1^2}
    F_\|+F_\perp+ \mathsf{P_1} Z_1}\bigg] }~,
\end{align}
where, $Z_1$ is defined as $Z_1=\sqrt{4 F_\|F_\perp-\tfrac{16}{9} A_{\text
    FB}^2}$ and the same sign in the remaining sign ambiguity must be
chosen for all $C_9$, $C_{10}$ and $\widetilde{\mathcal{G}}_\|$.  We
note that our solutions of the Wilson coefficients depend explicitly
on the assumption that $A_{\text FB}\neq 0$, hence, Wilson
coefficients can be determined at any $q^2$ except at the zero
crossing of $A_{\text FB}$. Since the Wilson coefficients are expected
to be real constants independent of $q^2$ away from the resonant
region, $Z_1$ must be real, implying the constraint
\begin{equation}
\label{real-Z1}
 F_\|F_\perp\geq \frac{4}{9} A_{\text FB}^2
\end{equation}
purely in terms of observables. The violation of this constraint will
be a clean signal of new physics.

\begin{figure}[!bt]
  \centering
\includegraphics*[width=1.65in]{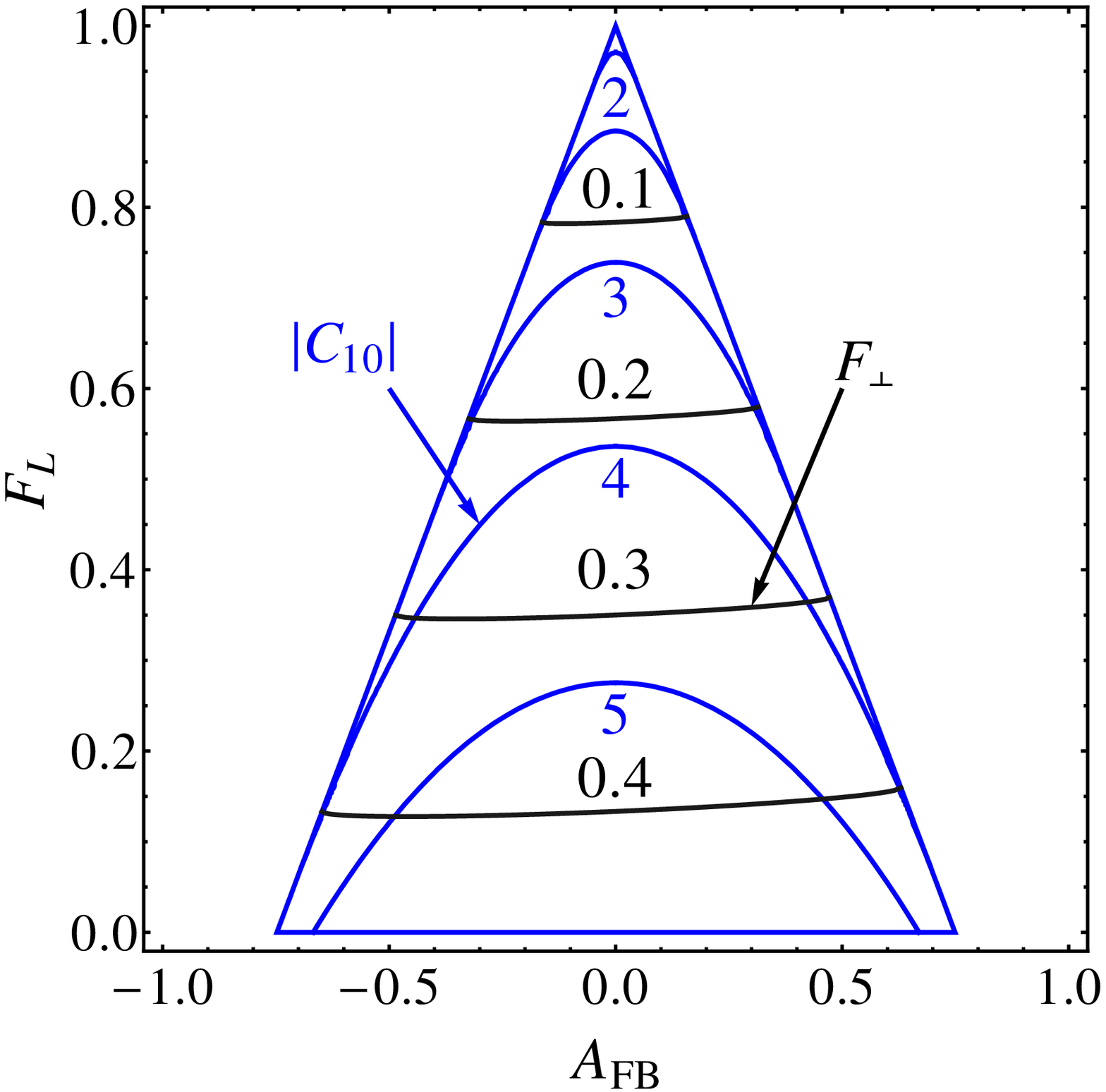}%
\includegraphics*[width=1.65in]{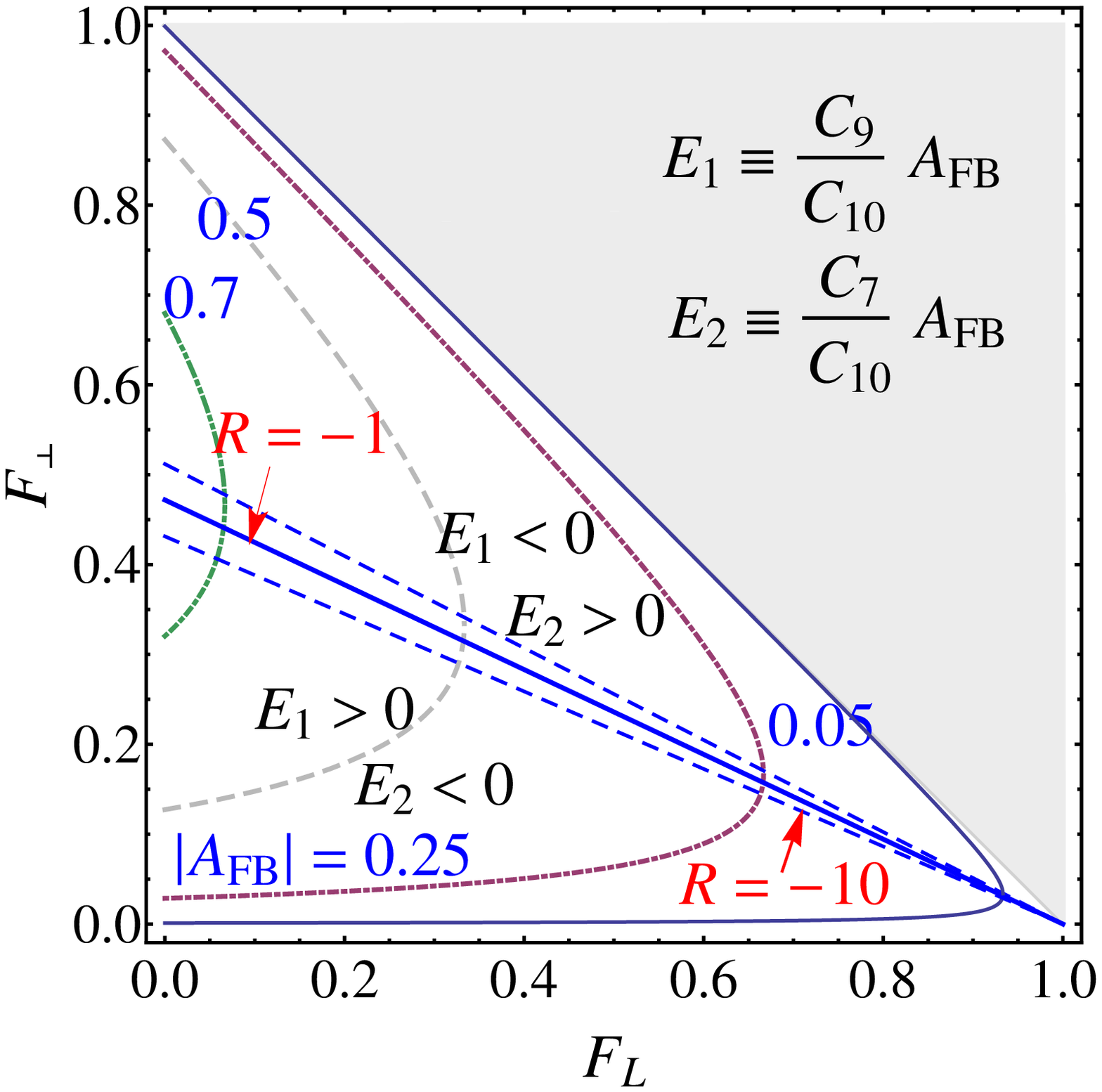}%
\caption{Left: The(blue) triangular region constrains $F_L - \AFB$
  parameter space. The solid (blue) lines correspond to $|C_{10}|$
  contours from Eq.~(\ref{eq:C10}) for $\Gamma_{\!\! f}=0.42\times
  10^{-7}$. Also plotted is $F_\perp$ for $R=-1$.  For
  LHCb~\cite{Aaij:2011aa} data with $1\,\gev^2\le q^2\le 6\,\gev^2$ we obtain
  $|C_{10}|=3.81\pm 0.58$ and $F_L=0.21\pm 0.05$.  Right: The
  constraints on $F_L$ and $F_\perp$ arising from
  Eq.~(\ref{eq:A_FB}). For $R=-1$ the allowed values lie on the
  diagonal solid (blue) line and for $R=-10$ between the two dashed
  lines. Also indicated is the $|\AFB|$ domain implied by $Z_1^2>0$.
  The shaded gray area is forbidden by
  $F_L+F_\perp+F_\parallel=1$. The $R=-1$ line approximately divides
  the domain into regions fixing the sign of $\AFB$ relative to
  $C_9/C_{10}$ and $C_7/C_{10}$ independent of $R$. }
   \label{fig:wedge}
 \end{figure}

Eq.~(\ref{eq:C10}) implies the following bound on the form factor $\mathsf{P_1}$
in terms of observables alone:
\begin{equation}
  \label{eq:7a}
  \mathsf{P_1^2}\lessgtr  \frac{4 F_\|F_\perp-\tfrac{16}{9} A_{\text
    FB}^2}{F_\|^2} \quad\forall\, F_\|F_\perp \lessgtr
 \frac{2}{7}\Big(\frac{4A_{\text{FB}}}{3}\Big)^2 
\end{equation}
The above bound is obtained by straight forward extremization of $\mathsf{P_1^2}$
with respect to all non-observables.  Eqs.~(\ref{eq:C9}) and (\ref{eq:C10}) result in the
following:
\begin{align}
  \label{eq:C9by10}
  &E_1\equiv \frac{C_9}{C_{10}}\AFB=\frac{3 (F_\|\mathsf{P_1} \mathsf{P'_1}-F_\perp) -(\mathsf{P_1}-\mathsf{P'_1})
    Z_1}{2(\mathsf{P_1}-\mathsf{P'_1})},
\end{align}
which can be inverted to express $A_{\text FB}$ in terms of $\mathsf{P_1}$, $\mathsf{P'_1}$
and $R$ as:
\begin{align}
  \label{eq:A_FB}
  \AFB=\frac{3 \left(R X-\sqrt{Y (\mathsf{P_1}-\mathsf{P'_1})^2(1+R^2)-X^2}\right)}{4
    (\mathsf{P_1}-\mathsf{P'_1}) \left(1+R^2\right)}
\end{align}
where, $X=2(F_\| \mathsf{P_1} \mathsf{P'_1}-F_\perp)$ and $Y=4 F_\|F_\perp$. 
Since $F_L$ has been measured and $F_L+F_\|+F_\perp=1$, all our conclusion 
throughout
the paper can be re-expressed in terms of just two helicity fractions
$F_L$ and $F_\perp$.  The usefulness of Eq. ~(\ref{eq:A_FB}) is shown
in Fig.~(\ref{fig:wedge}), where we have depicted the allowed
parameter space consistent with real $\AFB$. The reader will note that
the rigorous constraint imposed on $F_\perp$ depending on the value of
$F_L$ indicated in the figure.

We now derive some useful relations that involve $C_7$ and are hence valid
only at the leading order. Eqs.~(\ref{eq:C10}) and (\ref{eq:C7}) can
be re-expressed in this limit as: 
\begin{align}
  \label{eq:C7by10}
  &E_2\equiv \frac{C_7^\eff}{C_{10}} \AFB=
  \frac{3}{2} \frac{{\mathcal
      F}_\|}{\mathcal{G}_\|}\frac{(\mathsf{P_1^2} F_{\|}-F_\perp)}{ (\mathsf{P_1}-\mathsf{P'_1})}~, 
\end{align}
We emphasize that $C_7^\eff/C_{10}$ is not as clean as $C_9/C_{10}$, which
is expressed in Eq.~(\ref{eq:C9by10}) in terms of observables and
ratio's of two form factors which are predicted exactly in heavy quark
effective theory. $C_7^\eff/C_{10}$ on the other-hand depends on
$\mathcal{F}_\|/\mathcal{G}_\|$ which in turn depends on the heavy
quark effective theory form factor $\xi_\perp$. It may nevertheless be
noted that the sign of $\mathcal{F}_\|/\mathcal{G}_\|$ is quite
accurately predicted to be negative~\cite{Das-Sinha-II}.

Eq.~(\ref{eq:C9by10})
together with Eq.~(\ref{eq:C7by10}) can be rewritten in a form,
\begin{equation}
  \label{eq:7}
  \frac{2}{3} E_2 \mathsf{P_1^{'\!'}}-\frac{4}{3}
  E_1 \mathsf{P_1}\! =\!(\mathsf{P_1^2} F_{\|}+F_\perp+ \mathsf{P_1}Z_1)\!>\!0,
\end{equation}
where
$\mathsf{P_1^{'\!'}}=(\mathcal{G}_\|/\mathcal{F}_\|)\,
(\mathsf{P_1}+\mathsf{P'_1})>0$ since each of 
$(\mathcal{G}_\|/\mathcal{F}_\|)$, $\mathsf{P_1}$ and $\mathsf{P'_1}$
are always negative; $(\mathsf{P_1^2} F_{\|}+F_\perp+
\mathsf{P_1}Z_1)$ is easily seen to be always 
positive by an (infinite) series expansion in $\AFB$ where every terms
is positive.  In SM, $C_7^\eff/C_{10} >0$ and $C_9/C_{10}<0$, hence the
sign of $E_2$ ($E_1$) will be same (opposite) to that observed for
$\AFB$. If for any $q^2$ we find $\AFB>0$, Eq.~(\ref{eq:7}) cannot be
satisfied unless the contribution from the $E_2$ term exceeds the
$E_1$ term, or the sign of the $E_2$ term is wrong in SM. In
the SM the $E_2$ term dominates at small $q^2$, hence, $\AFB$ must be
positive at small $q^2$ to be consistent with SM.  If $\AFB<0$ is
observed for all $q^2$ i.e. no zero crossing of $\AFB$ is seen, one can
convincingly conclude that $C_7^\eff/C_{10}<0$ in contradiction with
SM. However, if zero crossing of $\AFB$ is confirmed with $\AFB>0$ at
small $q^2$ it is possible to conclude that the signs $C_7^\eff/C_{10}>0$
and $C_9^\eff/C_{10}<0$ are in conformity with SM.

The relations derived above depend only on three observables: $F_\|$,
$F_\perp$ and $\AFB$. Similar relations can be derived
using the three other nonzero observables $F_L$, $A_4$ and
$A_5$:
It is easy to derive these relations which are identical except for
the replacements: $F_\|\to F_L$, $\AFB \to \sqrt{2}A_5$, $\mathcal{F}_\| \to
\mathcal{F}_0$ and $\mathcal{G}_\| \to \mathcal{G}_0$ in
Eqs.~(\ref{eq:C9}), (\ref{eq:C10}) and (\ref{eq:C7}). We obtain
$C_9/C_{10}$ and $C_7^\eff/C_{10}$ ratios as
\begin{align}
\label{eq2:C9byC10}
\frac{C_9}{C_{10}}\sqrt{2}A_5=&\frac{3(F_L
  \mathsf{P_2}\mathsf{P_2'}-F_\perp)-(\mathsf{P_2}-\mathsf{P_2'})Z_2}{2
  (\mathsf{P_2}-\mathsf{P_2'})}~,\\
\label{eq2:C7byC10}
\frac{C_7^\eff}{C_{10}}\sqrt{2}A_5=&\frac{3}{2}\frac{\mathcal{F}_0}{\mathcal{G}_0}
\frac{(\mathsf{P_2^2}F_L-F_\perp)}{(\mathsf{P_2}-\mathsf{P_2'})}
\end{align}
where $Z_2=\sqrt{4 F_L F_\perp-\frac{32}{9} A_{5}^2}$ and we have
defined  $\mathsf{P_2}\equiv \mathcal{F}_\perp/\mathcal{F}_0$ and $\mathsf{P_2'}\equiv \mathcal{G}_\perp/\mathcal{G}_0$.

It is straight forward to derive relations analogous to
Eqs.~(\ref{real-Z1}), (\ref{eq:7a}), (\ref{eq:A_FB}) and ( \ref{eq:7})
in terms of $F_L$, $F_\perp$, $A_5$, $\mathsf{P_2}$ and $\mathsf{P_2'}$.
We present here only one of the interesting
possible relation
\begin{align}
  \label{eq:realZ3}
  4 (1-F_\perp) F_\perp\geq \frac{16}{9}(\AFB^2+2 A_5^2),
\end{align}
which bounds $\AFB^2$ and $A_5^2$ in terms of $F_\perp$ alone, and
is obtained by generalizing the constraints obtained from $Z_1>0$ and
$Z_2>0$.
$\mathsf{P_2}$ and $\mathsf{P_2'}$ are obtained in terms of observables and $\mathsf{P_1}$, $\mathsf{P'_1}$ by
comparing the expressions for $C_9/C_{10}$ in Eqs.~(\ref{eq:C9by10})
and (\ref{eq2:C9byC10}) and for $C_7/C_{10}$ in Eqs.~(\ref{eq:C7by10})
and (\ref{eq2:C7byC10}). We find,
\begin{equation}
  \label{eq:P2}
\mathsf{P_2}=\frac{2\mathsf{P_1}\AFB F_\perp}{\sqrt{2}A_5(2F_\perp+Z_1\mathsf{P_1})-Z_2\mathsf{P_1} \AFB}
\end{equation}
The expression for $\mathsf{P_2'}$ is some what more complicated and hence we
refrain from presenting it here. 
We note that $\mathsf{P_2}$, $\mathsf{P_2'}$ depend on form factors
$\xi_\perp$ and $\xi_\|$, and are hence not regarded as clean
parameters. Nevertheless, we have shown that it is possible to
estimate both $\mathsf{P_2}$ and $\mathsf{P_2'}$ in terms of
observable and $\mathsf{P_1}$ and $\mathsf{P'_1}$ which are clean and
independent of the form factors. An astute reader will realize that
the expressions for $\mathsf{P_2}$ and $\mathsf{P_2'}$ are valid
beyond leading order. There is yet one more set of relations for
$C_7/C_{10}$ and $C_9/C_{10}$ that can be derived involving $A_4$ with
the substitution: $F_\|\to F_0+F_\|+\sqrt{2}\pi A_4$, $\AFB\to
\AFB+\sqrt{2}A_5$, $\mathcal{F}_\| \to \mathcal{F}_\|+\mathcal{F}_0$,
$\mathcal{G}_\| \to \mathcal{G}_\|+\mathcal{G}_0$, $\mathsf{P_1}\to
\mathsf{P_3}\equiv \mathsf{P_1}
\mathsf{P_2}/(\mathsf{P_1}+\mathsf{P_2})$ and $\mathsf{P'_1}\to
\mathsf{P_3'}\equiv \mathsf{P'_1}
\mathsf{P_2'}/(\mathsf{P'_1}+\mathsf{P_2'})$.  Akin to the relation
for $\mathsf{P_2}$, $\mathsf{P_3}$ can also be obtained in terms of
$\mathsf{P_1}$ and observables.
However, $\mathsf{P_3}$ is related to $\mathsf{P_1}$ and $\mathsf{P_2}$. This results in an
relation for $A_4$:
\begin{equation*}
  A_4=
\frac{8 A_5 \AFB}{9 \pi F_\perp}+\sqrt{2}\,\frac{\sqrt{F_LF_\perp
       -\frac{8}{9} A_5^2}\sqrt{F_\|F_\perp -\frac{4}{9}\AFB^2}}{\pi F_\perp}.
\end{equation*}
 We have only briefly outlined the derivation of the
relation for $A_4$ that depends on observables alone.  Details of this
derivation and a complete set of constraints can be found in
Ref.~\cite{Das-Sinha-II}.

It is not surprising that several constraints exist between
observables in SM and that one is even able to predict the unmeasured
observables.  We have six observables expressed in terms of real
transversity amplitudes ${\cal A}_\lambda^{L,R}$, involving eight
parameters. Since $\mathsf{P_1}$, $\mathsf{P'_1}$ and $R$ are reliably
estimated in the SM, we have judiciously chosen them as additional
inputs to predict observables. Since only $\Gf$, $F_L$ and $\AFB$ are
measured, we can only predict $F_\perp$. 
However, with the additional assumption of $A_5$ that can be easily
measured we can predict $A_4$ purely in terms of observables.

We conclude that angular analysis in $B\to \kstar \ell^+\ell^-$, can
provide several tests of the Standard Model that are independent of
universal form factors. We find that the allowed parameter space for
observables is very tightly constrained, enabling significant tests of
the standard model. We also derived relations between observables and
form factors that are independent of Wilson coefficients and provide
interesting test of our understanding of hadronic effects contributing
to this decay. We also find that within SM there exists a relation
involving only observables, the violation of which would signal New
Physics. The several relations derived by us will not only test our
understanding of hadronic parameters but also provide unambiguous
indications if New Physics contributes to these decays.

\end{document}